\documentclass[aps,prresearch,reprint,amsmath,amssymb,superscriptaddress,
               floatfix]{revtex4-2}

\usepackage{graphicx}
\newcommand{\logB}{\ln\mathrm{B}}
\usepackage[hidelinks]{hyperref}

\begin{document}

\title{The number of communities in real networks grows as a power law
of their size}

\author{Alexei Vazquez}
\email[Contact author: ]{alexei.vazquez@gmail.com}
\affiliation{Nodes \& Links Ltd, Salisbury House, Station Road,
Cambridge, CB1 2LA, UK}

\date{\today}

\begin{abstract}
Most networks, from friendships to the internet, break up into
communities: groups of nodes more densely connected to one another than to the
rest. A basic question has remained open: how many communities should a network
have, and how does that number grow with size? Simple models bracket the
possibilities. In a small-world network built from dense modules---a caveman
graph of cliques---each module is a community, so the number grows in proportion
to size, as $n$. In self-similar, fractal networks it grows far more slowly, as
the square root. Real networks could lie anywhere between. Here we measure the
number of communities across close to a hundred real networks spanning four
orders of magnitude, from a $34$-member club to millions of nodes, and within
individual systems followed over decades as they grow. We
find a power law with an exponent between one half and two thirds---above the
fractal value and below the linear limit. Simple models of network growth by
local rules reproduce the same super-fractal exponent, a stable and measurable
property of network organization. In short, network evolution herds a system's
components into clusters, whose number grows as a power law of size.
\end{abstract}

\maketitle

Systems as different as social groups, power grids, protein interactions and
scientific collaborations are described as networks of nodes joined by
links\cite{watts1998,barabasi1999,boccaletti2006}. A striking regularity of
these networks is that they are not homogeneous: their nodes fall into
communities, tightly knit groups that often correspond to functional units such
as social circles, metabolic pathways or research fields\cite{girvan2002}.
Detecting communities, and measuring how good a division is, has become one of
the central tools of network science\cite{newman2004,fortunato2010}. Yet a deceptively simple
question is rarely asked: given a network of $n$ nodes, into how many communities
does it naturally divide?

\begin{figure}[t]
  \centering
  \includegraphics[width=\columnwidth]{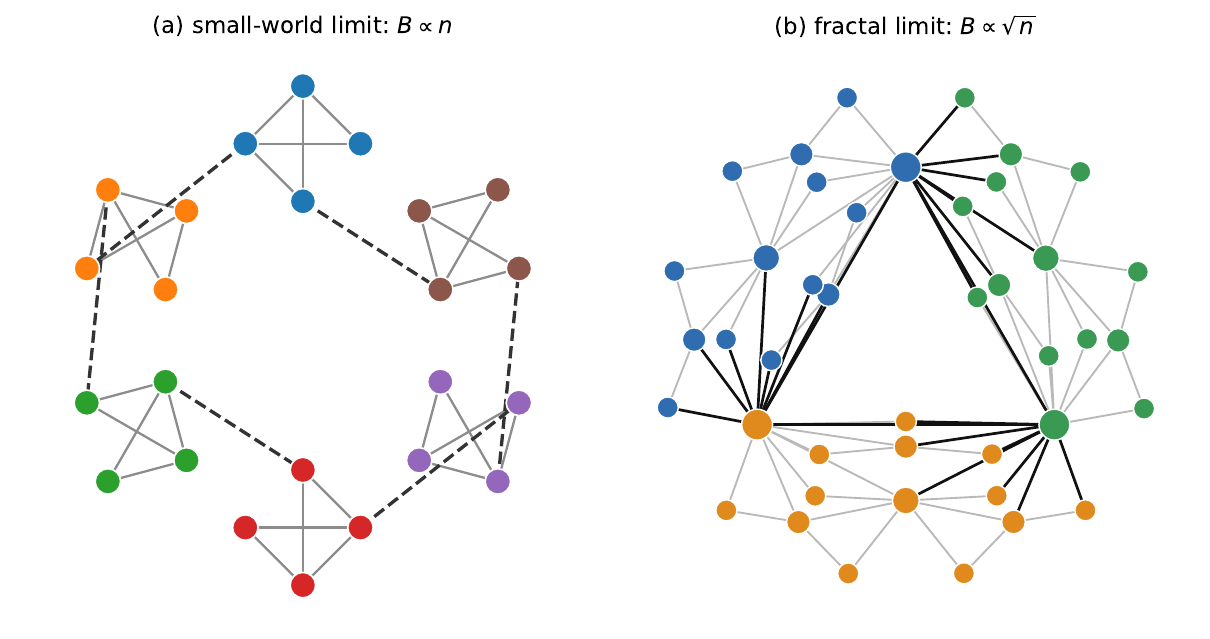}
  \caption{\textbf{Two model limits for the growth of communities}, with
  communities color coded. \textbf{(a)} The small-world limit: a connected
  caveman graph of $\ell=6$ cliques of $k=4$ nodes ($n=24$), each clique its own
  community (gray edges within a clique, dashed edges the bridges joining them into
  a ring). Above a critical clique size $k_c=4$ the Bayesian evidence prefers one
  community per clique (Appendix~\ref{app:bound}), so the number of communities equals the number
  of cliques, $B=n/k$, growing linearly ($\beta=1$). \textbf{(b)} The fractal
  limit: the pseudofractal web of Dorogovtsev, Goltsev and Mendes\cite{dgm2002} at
  generation $t=3$ ($n=42$), colored by its symmetric three-community partition (one descent
  branch plus its seed hub each; dark edges mark the seams between communities),
  with node size proportional to degree (the three large hubs are the seed
  triangle). For this self-similar network the number of communities grows as the
  square root, $B\propto\sqrt{n}$ ($\beta=\tfrac12$)\cite{pseudofractal}. Real
  networks fall between these limits, at an exponent between one half and two
  thirds.}
  \label{fig:caveman}
\end{figure}

The answer is not obvious, but two kinds of idealized model bracket it. At one
extreme are small-world networks built from dense modules. The clearest is the
caveman graph of Watts\cite{caveman}, a ring of tightly knit cliques
(Fig.~\ref{fig:caveman}a): each clique is unambiguously its own community, so the
number of communities is proportional to the number of nodes, $B\propto n$, and
communities keep a fixed typical size (Appendix~\ref{app:bound}). At the other extreme are
self-similar, or fractal, networks, such as the pseudofractal web of Dorogovtsev,
Goltsev and Mendes\cite{dgm2002} (Fig.~\ref{fig:caveman}b). When such a
network is described by a
statistical model that rewards a good fit but penalizes needless complexity, the
best description balances too few communities, which blur real structure, against
too many, which merely fit noise; for fractal networks this balance can be worked
out exactly and falls at a number of communities that scales as the square root,
$B\propto\sqrt{n}$, each community itself containing about $\sqrt{n}$
nodes\cite{pseudofractal,rgcrossing}. The same square-root scale appears, from a
different direction, in the resolution limit of community
detection\cite{fortunato2007,peixoto2013}. These two mechanisms set a range for
the growth of communities, from the fractal exponent $\tfrac12$ to the fully
modular exponent $1$ in $B\propto n^{\beta}$. Where real networks fall within this
range is an empirical question.

To answer this we collected close to a hundred undirected networks from a public
repository\cite{netzschleuder}, chosen to span the widest possible range of
sizes, from Zachary's karate club ($n=34$) to networks of more than a million
nodes. They come from social, biological, infrastructural and collaboration
settings, with no common design. For each network we counted its communities with
the degree-corrected stochastic block model, a Bayesian method that infers the
number of groups directly from the data, without it being fixed in
advance\cite{peixoto2013,peixoto2019}. Because the inference involves a random
search, we repeated it several times for every network and report the average and
its spread.

\begin{figure*}[t]
  \centering
  \includegraphics[width=\textwidth]{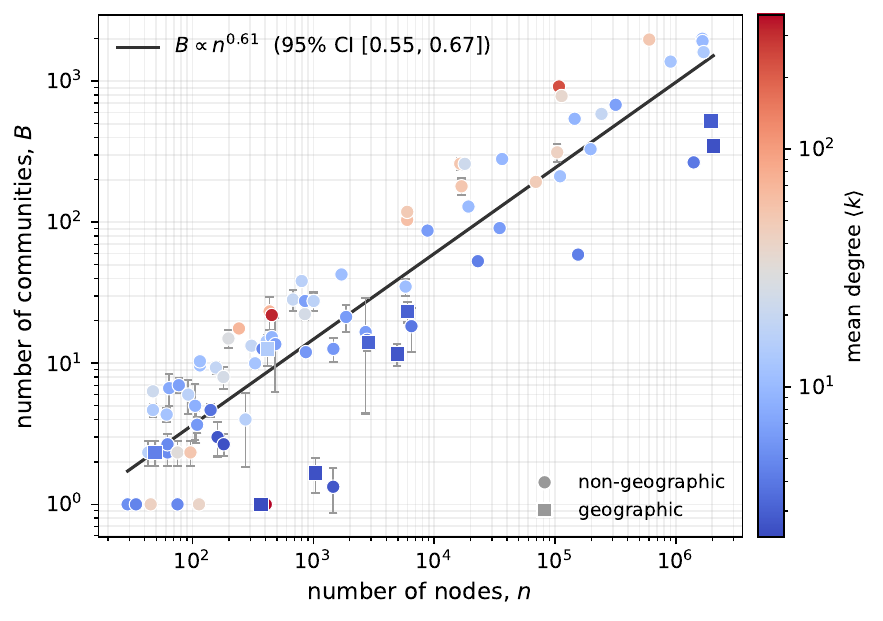}
  \caption{\textbf{The number of communities grows as a power law of network
  size.} Number of communities $B$ versus number of nodes $n$ for close to a
  hundred real networks (log--log axes). Points are the mean over independent
  inferences, error bars one standard deviation, colored by mean degree
  $\langle k\rangle$ (color bar; dense red, sparse blue), with geographic networks
  drawn as squares. The solid line is a power-law fit; the exponent
  $\beta\approx0.61$ (bootstrap $95\%$ CI in the legend) lies well above the
  fractal value $1/2$.}
  \label{fig:data}
\end{figure*}

Figure~\ref{fig:data} shows the result. On logarithmic axes the number of
communities $B$ follows a clean straight line over more than four decades in
size: a power law $B\propto n^{\beta}$. The fitted exponent is $\beta\approx0.61$,
with a bootstrap $95\%$ confidence interval that lies well above one half and
excludes it. The number of communities therefore grows as a power law of size,
but faster than the $\sqrt{n}$ predicted for fractal networks. The karate club
resolves into a single group, the football schedule into ten, the internet into
several dozen, the largest collaboration and social networks into hundreds; a
single exponent connects them across more than four orders of magnitude, even
though the networks share no common origin, generative rule or scale.

The scatter around the line is informative rather than troubling, and tracks the
network density, with which the points in Fig.~\ref{fig:data} are colored. Denser
networks (red), such as social ego-networks, sit somewhat above the line, and
sparse, spatially embedded ones (blue), such as a road network, fall below it: at
a fixed number of nodes a denser web carries more information and supports finer
division. Across the sample this is significant: networks above the line have a
higher mean degree than those below (Mann--Whitney $p<10^{-6}$), and the residual
from the fit correlates positively with mean degree (Spearman $\rho\approx0.6$,
$p<10^{-10}$). Geographic, spatially embedded networks (squares in
Fig.~\ref{fig:data})---roads, power grids and transport---are the sparsest and lie
furthest below the line: both their residual and their mean degree are
significantly lower than those of the non-geographic networks (Mann--Whitney
$p<10^{-3}$ and $p<10^{-4}$). Their position below the line is thus accounted for
by their low density rather than by geography as such. The
spread between repeated inferences, shown by the error bars, is small for most
networks and confirms that the count is a stable property, not an artifact of the
algorithm. What survives all of this variation is the trend: the number of
communities grows as a power law of size, with an exponent well above the fractal
one half.

\begin{figure*}[t]
  \centering
  \includegraphics[width=\textwidth]{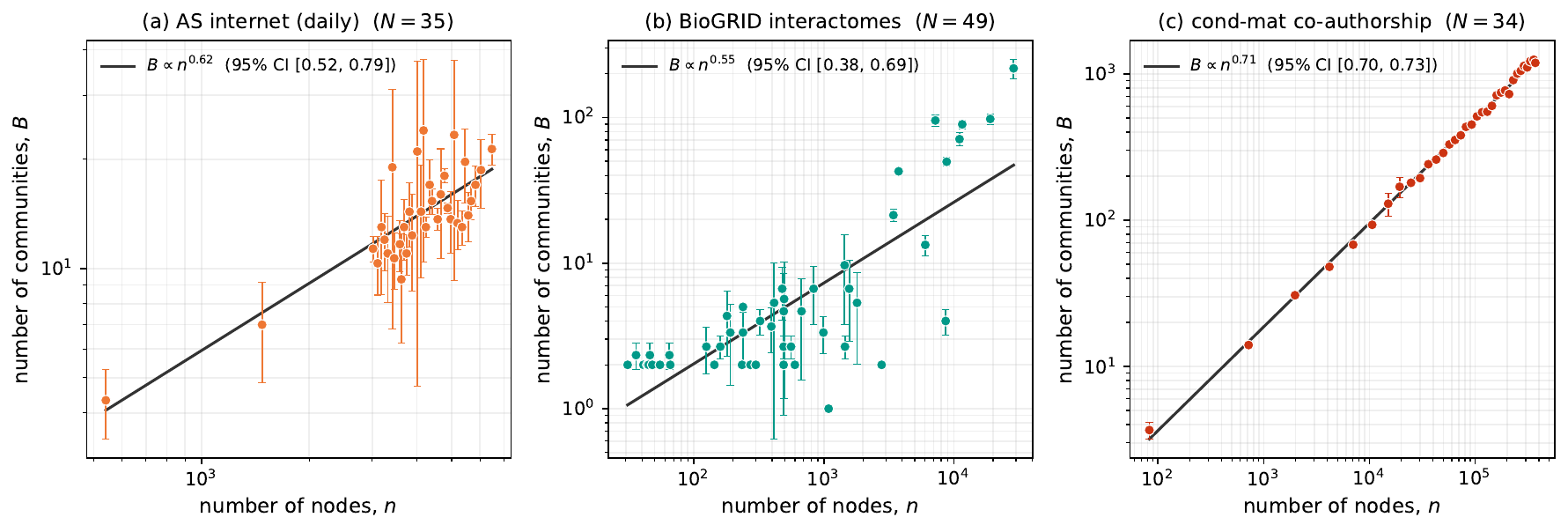}
  \caption{\textbf{The power law holds within individual growing systems.} Number
  of communities $B$ versus number of nodes $n$ (log--log, error bars one standard
  deviation) as three specific systems grow: (left) the internet at the autonomous
  system level from daily snapshots; (center) BioGRID physical protein
  interactomes across organisms; (right) the condensed-matter co-authorship
  network accumulated year by year, 1992--2026. Solid lines are power-law fits
  with the exponent and its bootstrap $95\%$ confidence interval; each system
  grows its community count as a power law of size, with exponents above the
  fractal $1/2$.}
  \label{fig:examples}
\end{figure*}

The cross-section of Fig.~\ref{fig:data} compares different networks. A sharper
test follows a \emph{single} system as it grows, so that the same rules build
every point. We did this for three systems tracked over time
(Fig.~\ref{fig:examples}): the internet at the level of autonomous systems, from
daily snapshots; the physical protein interactomes of many organisms; and the
co-authorship network of the condensed-matter physics literature, accumulated
year by year over three decades to nearly $400{,}000$ authors. Each system,
grown by its own mechanism, again produces a power law. The co-authorship
network, the cleanest and widest in range, gives an exponent $\beta\approx0.71$
with a tight confidence interval; the internet gives $\approx0.62$ and the
interactomes $\approx0.55$. Within individual growing systems, as across the
cross-section, the number of communities grows as a power law with an exponent
between one half and two thirds ($0.55$ to $0.71$ across systems), well above the
fractal one half.

\begin{figure*}[t]
  \centering
  \includegraphics[width=\textwidth]{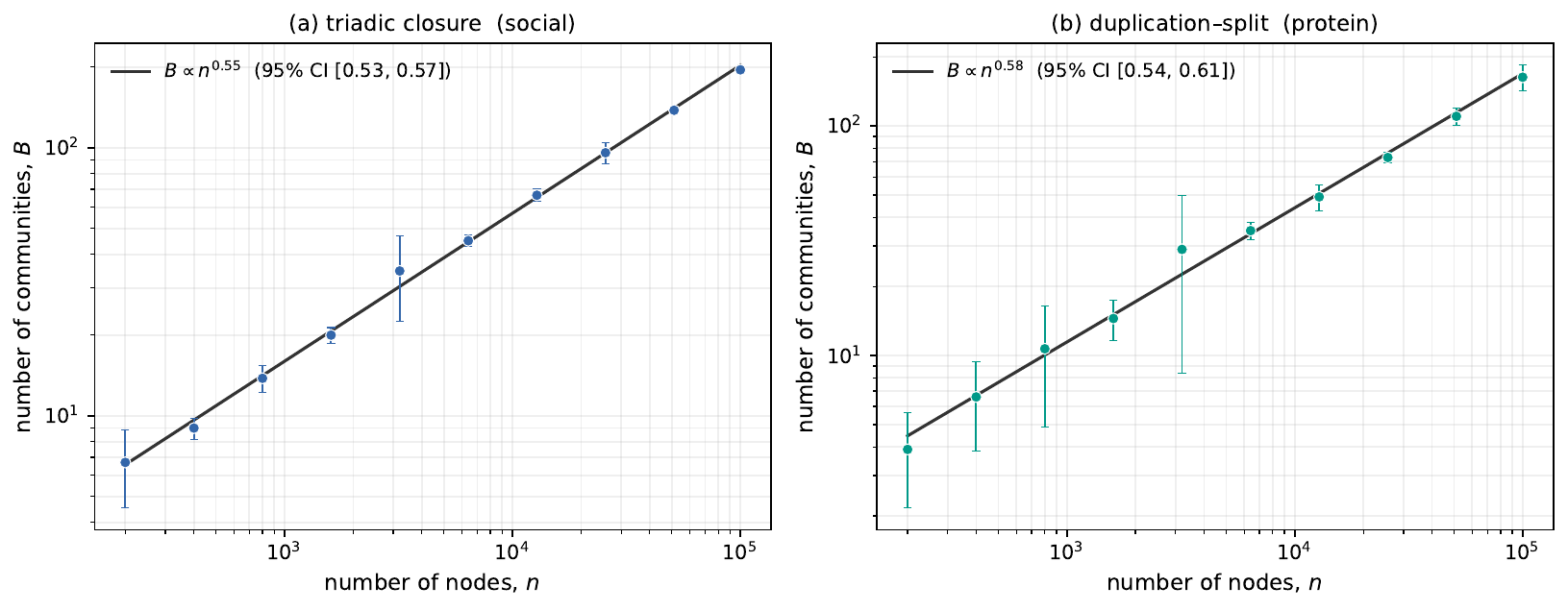}
  \caption{\textbf{Simple growth models reproduce the power law.} Number of
  communities $B$ versus number of nodes $n$ (log--log, mean $\pm$ standard
  deviation over ten realizations) for two generative models grown to $10^5$
  nodes: (left) triadic closure, a social-network model; (right) duplication--split,
  a protein-network model. Solid lines are power-law fits, with the exponent and
  its bootstrap $95\%$ confidence interval given in each panel. Both models follow
  a power law with a super-fractal exponent between one half and one.}
  \label{fig:models}
\end{figure*}

A further, mechanistic line of evidence comes from generative models. If the
power law reflects a basic feature of how networks grow, simple growth rules
should reproduce it\cite{vazquez03local,localrules}. We tested two minimal models (Fig.~\ref{fig:models}). In
\emph{triadic closure}, a stylized model of social networks, each new node
attaches to a random node and to one of its neighbors, closing a triangle; this
rule was introduced as a one-step network walk\cite{vazquez2001} and, in the
community context, as triadic closure\cite{bianconi2014}. In
\emph{duplication--split}, a stylized model of protein-interaction networks
evolving through gene
duplication\cite{vazquez03dup,pastor-satorras03,chung03}, each new node either
copies the links of a random node or takes some of them over. Grown to $10^5$ nodes and analyzed identically to the
data, both models produce clean power laws, with exponents $\beta\approx0.55$
(triadic closure) and $0.59$ (duplication--split), in both cases with a $95\%$
confidence interval that lies above the fractal one half. Local growth rules thus
generate the power law by themselves, and place its exponent in the same
super-fractal range, between one half and one, seen in the real networks.

This power law has a simple and general meaning. The natural level at which a
network can be described is not fixed, but grows with the network. A small network
is well captured by a handful of groups; a much larger one is best described by
more groups, though still far fewer than the number of nodes. With an exponent
$\tfrac12<\beta<\tfrac23$, the number of communities grows as $n^{\beta}$ and
their typical size as $n^{1-\beta}$, so that both grow without bound but the
communities remain a vanishing fraction of the whole. One cannot summarize an
arbitrarily large network with a fixed number of parts, nor split it into ever
more without limit; the exponent sets the balance. That real networks divide more
finely than the fractal value $\beta=\tfrac12$ means they carry more resolvable
community structure than self-similar models predict.

The measured exponent falls squarely between the two model limits introduced at
the outset. The fully modular limit is realized exactly by the caveman graph
(Fig.~\ref{fig:caveman}): above a critical clique size $k_c=4$ the Bayesian
evidence prefers one community per clique over no partition at all, so the number
of communities equals the number of cliques, $B=n/k$, growing \emph{linearly}
($\beta=1$; Appendix~\ref{app:bound}). The fractal limit $\beta=\tfrac12$ sits at the other end.
Real networks lie between, at an exponent between one half and two thirds: more
finely divided than fractal self-similar models, but far from the fully modular
extreme. Where a
network falls between one half and one is thus a substantive, measurable fact
about how it is organized\cite{rgcrossing,localrules}, not a mathematical
inevitability.

That a power law with a single super-fractal exponent should hold across social,
biological and technological networks, and within individual systems as they
grow, suggests it reflects a basic constraint of network organization rather than
the details of any one system, in the same spirit as the small-world and
scale-free regularities found earlier\cite{watts1998,barabasi1999}. The
exponent one half is a genuine prediction of self-similar, fractal
networks\cite{pseudofractal,rgcrossing}, related to the resolution limit of
community detection\cite{fortunato2007,peixoto2013}; that real networks
systematically exceed it shows they are not merely self-similar but carry extra,
finer community structure. That communities should appear at all, and multiply as
a network grows, need not be attributed to any special property of the nodes: it
is an inevitable outcome of local growth rules, since networks assembled by simple
local mechanisms acquire community structure with near-certainty, whereas their
randomized counterparts do not\cite{localrules}. The exponent gives a baseline
against which unusually coarse or unusually fine organization can be recognized,
and a natural scale at which to compare networks of very different sizes; its
consistent value above one half, and short of one, is itself the finding.
Ultimately, the evolution of a network herds its components into clusters, and
their number increases as a power law of the system's size.

\section*{Data availability}

\noindent All networks analyzed here are publicly available from the Netzschleuder
repository\cite{netzschleuder}. The computed community counts are provided in the
code repository below.

\section*{Code availability}

\noindent The code that downloads the networks, runs the inference and produces
the figures is available at
\url{https://github.com/av2atgh/ramsey_netcom/sqrt}.

\begin{acknowledgments}
Nodes \& Links Ltd provided support in the form of salary for Alexei Vazquez,
but did not have any additional role in the conceptualization of the study,
analysis, decision to publish, or preparation of the manuscript. The
code and text were prepared with the assistance of Claude Opus~4.8, an AI
assistant developed by Anthropic.
\end{acknowledgments}

\section*{Author contributions}

\noindent A.V. designed and performed the research, analyzed the data and wrote
the manuscript. A.V.\ directed the use of the AI assistant, verified all results,
and takes full responsibility for the content.

\section*{Competing interests}

\noindent The author declares no competing interests.

\appendix

\section{Methods}
\label{app:methods}

\noindent\textit{Data.} Networks were obtained from the Netzschleuder
repository\cite{netzschleuder} through the \texttt{graph-tool} interface. We
used only undirected, non-bipartite networks, chosen to spread over the available
range of sizes. Each network was reduced to its largest connected component and
made simple (parallel edges and self-loops removed).

\noindent\textit{Community detection.} We inferred communities with the
degree-corrected stochastic block model by minimizing its description length
(\texttt{graph-tool}'s \texttt{minimize\_blockmodel\_dl})\cite{peixoto2013,%
peixoto2019}, which selects the number of communities automatically. The number
of communities is the number of non-empty groups in the inferred partition.
Because the optimization is stochastic, we ran it three times per network and
report the mean and standard deviation. The random seed was fixed so that the
figure is exactly reproducible.

\noindent\textit{Growing systems.} Three systems were followed as they grow
(Fig.~\ref{fig:examples}). The internet is the SNAP Autonomous Systems AS-733
collection, one graph per daily peering snapshot (subsampled evenly in time). The
interactomes are the physical protein--protein interaction networks of BioGRID,
one per organism. The co-authorship network is built from arXiv condensed-matter
metadata, accumulating all authors and their collaborations up to each year
$1992$--$2026$. Each snapshot was reduced to its largest connected component and
its communities counted exactly as above.

\noindent\textit{Generative models.} Two growth models were simulated
(Fig.~\ref{fig:models}). Triadic closure: starting from an edge, each new node
picks a uniformly random node, then one uniformly random neighbor of it, and
links to both. Duplication--split (rate $q=0.3$): each new node picks a random
node $j$ and, with probability $q$, copies all of $j$'s links (duplication),
otherwise takes over one of them (split). For each model and each of ten target
sizes from $200$ to $10^5$ we generated ten realizations, kept the largest
connected component, and counted communities with one SBM run each. The exponent
was obtained as for the data, by least squares of $\log B$ on $\log n$.

\noindent\textit{Fit.} Exponents $\beta$ are ordinary least-squares fits of
$\log B$ on $\log n$; the $95\%$ confidence interval is obtained by bootstrap over
networks (resampling the points with replacement and refitting, $4000$ times). To
test the effect of density we split the networks by the sign of their residual
from the fit (above or below the line) and compared their mean degree $2E/n$ with
a one-sided Mann--Whitney test, and correlated the residual with the mean degree
by Spearman's rank correlation.

\section{A power-law upper bound from the caveman graph}
\label{app:bound}

\noindent To show that linear growth,
$\beta=1$, is attainable and bounds the exponent from above, consider the
connected caveman graph $\mathrm{CC}(\ell,k)$, introduced by Watts\cite{caveman}.
One starts from $\ell$ separate cliques, or ``caves'', each a complete graph on
$k$ nodes in which everyone is connected to everyone else. These cliques are then wired into a ring by a minimal
modification: in each clique a single edge $(u,v)$ is deleted, and one of its
now-freed endpoints, say $v$, is instead connected to a node in the neighboring
clique. Every clique thus gives up exactly one of its internal links and gains
one link to the next clique, and following these bridges from clique to clique
traces out a closed ring of all $\ell$ cliques. The resulting network has
$n=\ell k$ nodes; each clique retains $\binom{k}{2}-1$ internal edges and is
attached to the rest of the network by only its two bridging edges, one to each
neighbor. A clique is therefore far denser inside than outside, increasingly so
as $k$ grows.

We decide whether this structure is worth resolving by exact Bayesian evidence,
with no stochastic search. Every node has essentially the same degree $k-1$, so
degree correction is unnecessary and a plain block model applies. We compare two
descriptions: one community per clique ($\ell$ communities), and no communities
at all (a single block). In the assortative model, links within a community occur
with probability $\theta_{\rm in}$ and links between communities with probability
$\theta_{\rm out}$, each given a $\mathrm{Beta}(a,a)$ prior and integrated out;
integrating $\theta^{e}(1-\theta)^{N-e}$ against the prior gives the Beta function
$\mathrm{B}(e+a,N-e+a)$. Writing $\logB(x,y)=\ln\Gamma(x)+\ln\Gamma(y)-
\ln\Gamma(x+y)$, the log-evidence ratio between the $\ell$-community and the
single-community descriptions is
\begin{align}
  \ln R = {}& \logB(e_{\rm in}+a,\,N_{\rm in}-e_{\rm in}+a) \notag\\
  & + \logB(e_{\rm out}+a,\,N_{\rm out}-e_{\rm out}+a) \notag\\
  & - \logB(m+a,\,N-m+a) - \logB(a,a) \notag\\
  & + \ln\frac{\Gamma(\ell a)\,\Gamma(k+a)^{\ell}}
              {\Gamma(a)^{\ell}\,\Gamma(n+\ell a)},
  \label{eq:lnR}
\end{align}
where $e_{\rm in}=\ell\bigl[\binom{k}{2}-1\bigr]$ intra-clique and
$e_{\rm out}=\ell$ bridge edges lie among $N_{\rm in}=\ell\binom{k}{2}$
within-community and $N_{\rm out}=\binom{n}{2}-N_{\rm in}$ between-community pairs,
$m=e_{\rm in}+e_{\rm out}$ is the total number of edges, and $N=\binom{n}{2}$. The
last term is the label prior. A positive $\ln R$ favors one community per clique.

Evaluating Eq.~\eqref{eq:lnR} with $a=1$, $\ln R$ is negative for every $\ell$
when $k\le3$: sparse triangles are never worth separating. For $k\ge4$ it turns
positive and grows with the number of cliques---for $k=4$, $\ln R=-0.2$, $+5.5$,
$+20.5$, $+56.4$, $+139.8$ at $\ell=2,4,8,16,32$, and larger still for $k>4$---so
the network is best described by one community per clique. The critical clique
size is therefore $k_c=4$, and above it the number of communities is
\begin{equation}
  B=\ell=\frac{n}{k},
\end{equation}
which grows \emph{linearly} with $n$ at fixed $k$ ($\beta=1$). This construction,
and the Ramsey-community analysis of networks grown by local rules more
generally\cite{rgcrossing,localrules}, realizes the linear extreme $B\propto n$.
Real networks fall between this bound and the fractal $\beta=\tfrac12$, at
$\tfrac12<\beta<\tfrac23$.

\end{document}